\documentclass[twocolumn,superscriptaddress,amsmath,amssymb,aps,pra]{revtex4}
\usepackage{amsmath,amsthm,amssymb}
\usepackage{latexsym}
\usepackage{amscd,graphicx}
\usepackage[titletoc]{appendix}
\usepackage{epsfig}
\usepackage{epstopdf}
\usepackage[normalem]{ulem}
\usepackage[dvipsnames]{xcolor}
\usepackage[colorlinks=true,linkcolor=red,citecolor=blue]{hyperref}

\begin{document}
	
	\title{Nonreciprocal Unconventional Photon Blockade with Kerr Magnons}
	
    \author{Xiao-Hong Fan}
    \affiliation{Department of Physics, Wenzhou University, Zhejiang 325035, China}
    
    \author{Yi-Ning Zhang}
    \affiliation{Department of Physics, Wenzhou University, Zhejiang 325035, China}

    \author{Jun-Po Yu}
	\affiliation{College of Resources and Environment, Yangtze University, Hubei 430100, China}
    
    \author{Ming-Yue Liu}
    \affiliation{Department of Physics, Wenzhou University, Zhejiang 325035, China}
    
    \author{Wen-Di He}
    \affiliation{Department of Physics, Wenzhou University, Zhejiang 325035, China}
    
    \author{Hai-Chao Li}
    \altaffiliation{hcl2007@foxmail.com}
    \affiliation{College of Physics and Electronic Science, Hubei Normal University, Huangshi 435002, China}
	
    \author{Wei Xiong}
    \altaffiliation{xiongweiphys@wzu.edu.cn}
    \affiliation{Department of Physics, Wenzhou University, Zhejiang 325035, China}
	
	\date{\today }

	\begin{abstract}
	
    Nonreciprocal devices, allowing to manipulate one-way signals, are crucial to quantum information processing and quantum network. Here we propose a nonlinear cavity-magnon system, consisting of a microwave cavity coupled to one or two yttrium-iron-garnet (YIG) spheres supporting magnons with Kerr nonlinearity, to investigate nonreciprocal unconventional photon blockade. The nonreciprocity originates from the direction-dependent Kerr effect, distinctly different from previous proposals with spinning cavities and dissipative couplings. For a single sphere case, nonreciprocal unconventional photon blockade can be realized by manipulating the nonreciprocal destructive interference between two active paths, via vary the Kerr coefficient from positive to negative, or vice versa. By optimizing the system parameters, the perfect and well tuned nonreciprocal unconventional photon blockade can be predicted. For the case of two spheres with opposite Kerr effects, only reciprocal unconventional photon blockade can be observed when two cavity-magnon coupling strengths Kerr strengths are symmetric. However, when coupling strengths or Kerr strengths become asymmetric, nonreciprocal unconventional photon blockade appears. This implies that two-sphere nonlinear cavity-magnon systems can be used to switch the transition between reciprocal and nonreciprocal unconventional photon blockades. Our study offers a potential platform for investigating nonreciprocal photon blockade effect in nonlinear cavity magnonics.

	\end{abstract}
	
	\maketitle
	
\section{introduction}
	
Recently, magnons, also known as spin waves, i.e., the collective spin excitations in ferro- and ferrimagnetic materials like yttrium-iron-garnet (YIG), have attracted considerable attention in condensed matter physics and quantum information science~\cite{Huebl-2013,Tabuchi-2014,Zhang-2014,Goryachev-2014,Wang-2016,Bhoi-2014,Bai-2015,ZhangD-2015,Li-2019,Hou}. Thanks to the high spin density and low damping of the YIG spheres, photons in microwave cavities can strongly couple to the magnons, giving rise to the field of cavity magnonics~\cite{Rameshti-2022,Lachance-2019,Yuan-2022}. Experimentally, sub-millimeter-scale YIG spheres and three-dimensional microwave cavities are frequently employed in cavity magnonics~\cite{Tabuchi-2014,Zhang-2014,Goryachev-2014,Wang-2016} for investigating numerous exotic phenomena~\cite{Rameshti-2022,Wang-2020}, such as magnon memory~\cite{ZhangX-2015}, spin current~\cite{Bai-2015,Bai-2017,Mukhopadhyay-2022}, entanglement~\cite{Yuanhy-2020,Mousolou-2021,ZhangZ-2019,Ren-2022}, dissipative coupling~\cite{Harder-2018,Grigoryan-2018,Wang-2019}, blockade~\cite{Xie-2020,Wang-2022,Jingjun}, non-Hermitian physics~\cite{ZhangD-2017,Harder-2017,Cao-2019,Zhao-2020,Zhanggq-2019}, dynamics of polaritons~\cite{Yao-2017}, spin interface~\cite{Tian-2023,Hei-2021}, state manipulation~\cite{Xu-2023,Yuan-2020,Sun-2021,Zhanggq-2023,Qi-2022}, microwave-optical transduction~\cite{Hisatomi-2016,Zhu-2020}. In addition, magnons can strongly interact with superconducting qubits, solid spins, and phonons, building diverse magnon-based hybrid quantum systems including qubit-magnon systems~\cite{Tabuchi-2015,Lachance-2020,Dobrovolskiy-2019,Wolski-2020,Xiong4-2022,neuman-2020,neuman1-2021,neuman2-2021,Skogvoll-2021,trifunovic-2013,Fukami-2021}, cavity magnomechanics~\cite{zhang-2016,Li-2018,Shen-2022,Li-2021}, optomechanical cavity magnonics~\cite{Chen,Proskurin-2018,Xiong-2023}, and cavity optomagnonics~\cite{Zhangx-2016,Osada-2016,Haigh-2016}.

With advanced experimental techniques, the magnon Kerr effect (the Kerr nonlinearity of magnons) steming from the magnetocrystalline anisotropy in the YIG~\cite{Zhangguoq-2019} has been demonstrated~\cite{Wang-2016,Wangyp-2018}, leading to the birth of nonlinear cavity magnonics~\cite{Zheng-2023}. Utilizing the magnon Kerr effect, multi-stability~\cite{Wang-2016,Shenrc-2021}, magnon entanglement~\cite{ZhangZ-2019}, strong spin-spin coupling~\cite{Xiong4-2022,Ji-2023,Skogvoll-2021}, superradiant phase transition~\cite{Liu-2023}, and sensitive detection~\cite{Zhanggq1-2023,Nair-2021} can be studied. Besides, the magnon Kerr effect can also be used to investigate nonreciprocical devices such as nonreciprocal entanglement~\cite{Chen,ChenJ}, nonreciprocal transimission~\cite{Kong-2019}, nonreciprocal excitation~\cite{XiongW23} and nonreciprocal higher-order sideband generation~\cite{Wangm-2021}. However, nonreciprocal single-photon blockade has not yet been revealed to date with the magnon Kerr effect, although various nonreciprocal devices have been widely investigated with spinning cavities~\cite{Huang-2018,Jiaoyf-2020, Zhang-2023,Jing-2021,Yao-2022} and dissipative coupling~\cite{Wangy-2022,Wang-2020}. Note that photon blockade is a purely quantum effect, which can be employed to achieve single-photon source devices and generate sub-Poissonian light~\cite{Shields-2007,Davidovich-1996,Imamoglu-1997}. At present, two classes of photon blockade are proposed: conventional~\cite{Birnbaum-2005,Zhou-2018,Hamsen-2017} and unconventional~\cite{Shen-2015,Zhou-2016,Imamoglu-2011,Zhou-2015,Flayac-2017,Snijders-2018} photon  blockade. The former is caused by strong anharmonicity of the eigenenergy spectrum, and the latter is formed by the destructive quantum interference in different transition paths under weak nonlinearity.


Here, we propose a scheme to realize a nonreciprocal unconventional single-photon blockade in a Kerr-modified cavity-magnon system, which consists of a microwave cavity coupled to one or two YIG spheres supporting Kerr magnons. The nonreciprocity is induced by the direction-dependent Kerr nonlinearity. Specifically, when the biased magnetic field is alinged along the crystal axis $[100]$ ($[110]$), the Kerr coefficient is positive (negative), which has been demonstrated experimentally~\cite{Wangyp-2018}. In the case of a single sphere in the cavity, only two interference passages are activated. By changing the Kerr coefficient from positive to negative (or vice versa), nonreciprocal destructive interference occurs, leading to the manifestation of nonreciprocal photon blockade. This phenomenon can be rigorously demonstrated through both analytical and numerical analyses, focusing on the equal-time second-order correlation function. When the system parameters are optimized, achieving the (ideal) perfect nonreciprocal photon blockade becomes feasible. Additionally, we illustrate that the degree of nonreciprocity can be finely tuned by manipulating system parameters, as evidenced by the study of the defined contrast ratio. When two spheres with opposite Kerr coefficients are considered, three active interference passages emerge. In the case of symmetrical coupling strengths and Kerr coefficients, two passages induced by magnon-photon couplings assume identical roles in destructively interfering with the passage created by the pumping field, thereby leading to reciprocal photon blockade. When two cavity-magnon coupling strengths or Kerr coefficients become asymmetric, two passages activated by the coupling strengths assume distinct roles in interfering with the pumping passage, resulting in nonreciprocal photon blockade, as evidenced by the corresponding contrast ratio. This indicates that two-sphere nonlinear cavity-magnon systems can be used to switch the transition between reciprocal and nonreciprocal photon blockades.  Our investigation opens up a promising avenue for engineering nonreciprocal devices in both single and multiple YIG spheres with magnon Kerr effect.

The rest paper is organized as follows: In Sec.~\ref{s2}, the
model is described, and the  effective non-Hermitian Hamiltonian
is given. Then we study the nonreciprocal photon blockade in a cavity including a single sphere in Sec.~\ref{s3}. In Sec.~\ref{s4}, we further study the nonreciprocal photon blockade in a cavity including two symmetric and asymmetric spheres. Finally, a conclusion is given in  Sec.~\ref{s6}.

     \begin{figure}
     	\includegraphics[scale=0.34]{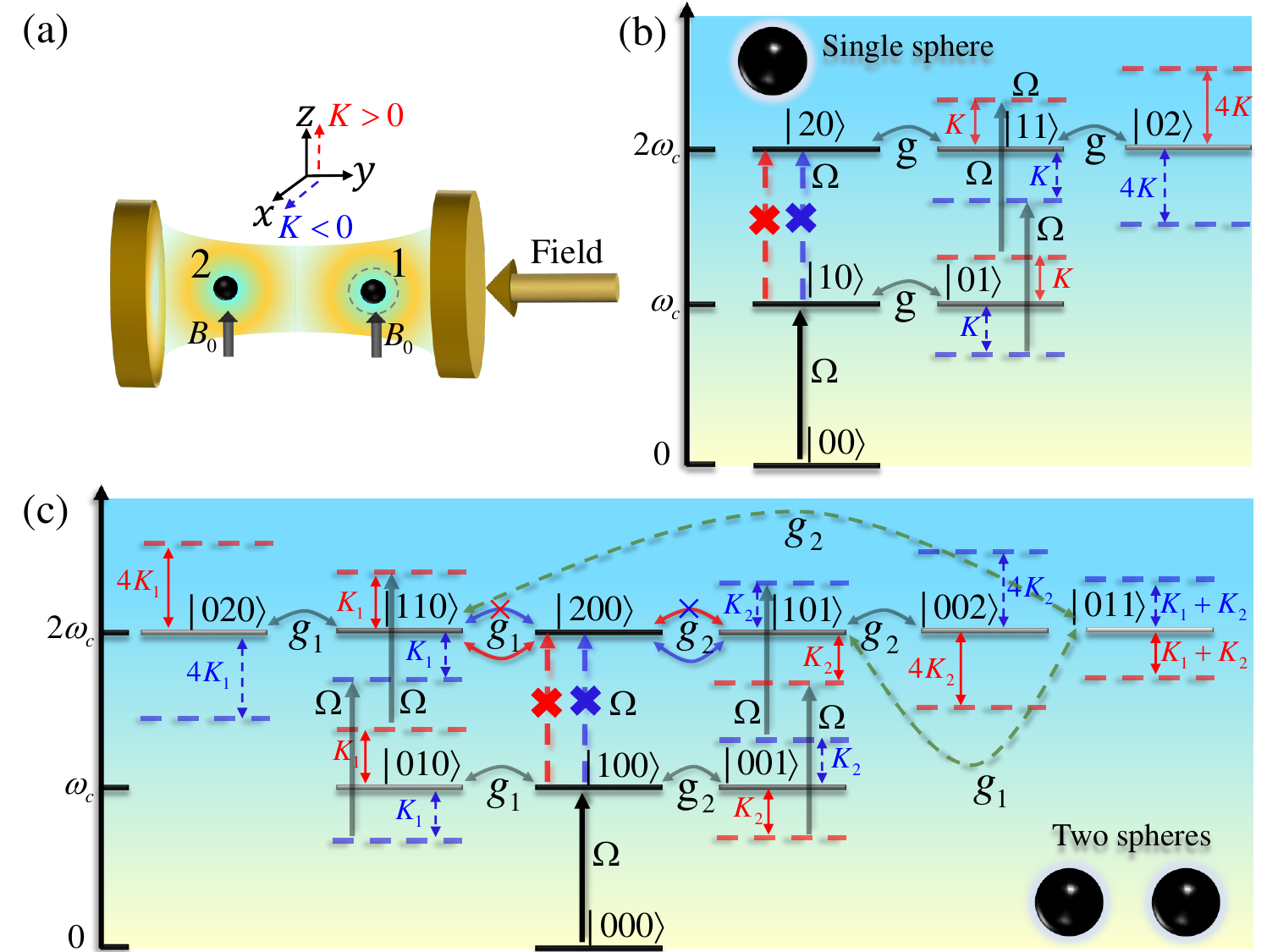}
     	\caption{(a) Schematic diagram of the proposed cavity-magnon system. It consists of one or two YIG spheres supporting Kerr magnons coupled to a pumped cavity. The YIG sphere(s) is (are) placed in a static magnetic	field $B_0$, along the crystallographic axis $[100]$ or $[110]$. Correspondingly, $K>0$ or $K<0$. (b) Energy level diagram of a single sphere coupled to a cavity and the corresponding excitation paths. (c) Energy level diagram of two spheres simultaneously coupled to a common cavity and the corresponding excitation paths.} \label{fig1}
     \end{figure}	
	\section{Model and Hamiltonian}\label{s2}

    We consider a nonlinear cavity magnonics consisting of one or two YIG spheres coupled to a microwave cavity [see Fig.~\ref{fig1}(a)], where the Kittel mode of the YIG sphere is used to support the Kerr magnons (i.e., magnons with the Kerr effect). Such the nonlinearity, arising from the magnetocrystallographic
    anisotropy, can be tuned by the direction of the magnetic field~\cite{Wangyp-2018,Zhanggq-2019}. Specifically,  the Kerr coefficient is positive (negative) when the magnetic field is aligned along with the crystallographic axis $[100]~([110])$ of the YIG sphere. For studying photon blockade effect, an additional pumping field with the frequency $\omega_p$ and the Rabi frequency $\Omega$ is imposed to the microwave cavity. The Hamiltonian of the proposed system can be written as (setting $\hbar=1$),
    
    \begin{align}
    	H_{\rm sys} =& \sum_{j=1,2}[{\omega _m}{m_j^\dag }m_j + g_j({m_j^\dag }c + {c^\dag }m_j) + K_j(m_j^\dag m_j)^2] \notag\\
    	&+{\omega _c}{c^\dag }c + \Omega (c{e^{i{\omega_p}t}} + {c^\dag }{e^{ - i{\omega_p}t}}),\label{eq1} 
    \end{align}
    where $\omega_{c(m)}$ is the resonance frequency of the photons (magnons) in the cavity (Kittel) mode, $g_j$ is the photon-magnon coupling strength and $K$ is the Kerr coefficient. The operators $c~(m_j)$ and $c^\dag~(m_j^\dag)$ are the annihilation and creation operators of the photons ($j$th magnon). In the rotating frame with respect to $\omega_p$, Eq.~(\ref{eq1}) reduces to
    \begin{align}
    H_{\rm rf}=& \sum_{j=1,2}[{\Delta_m}{m_j^\dag }m_j + g_j({m_j^\dag }c + {c^\dag }m_j) + K_j(m_j^\dag m_j)^2] \notag\\
    &+{\Delta_c}{c^\dag }c + \Omega (c + {c^\dag }),\label{eq2} 
    \end{align}
    where $\Delta_{c(m)} = \omega_{c(m)} - \omega_p$ is the frequency detuning of the photons (magnons) from the pumping field.
  
   By further taking the dissipations of the system into account and neglecting the quantum jump terms, the effective non-Hermitian Hamiltonian of the system is 
    \begin{align}
    	H_{\rm eff} = H_{\rm rf}-i\frac{\kappa_c}{2}c^\dag c-i\sum_{j=1,2}\frac{\kappa_m}{2}m_j^\dag m_j,\label{eq3} 
    \end{align}
where $\kappa_c$ and $\kappa_m$ are the decay rates of the photons and magnons, respectively.
   
    \section{Nonreciprocal photon blockade with A Single Sphere}\label{s3}

    In this section, we investigate the photon blockade in the proposed system consisting of a single YIG sphere coupled to the cavity, i.e., $j=1$ in Eq.~(\ref{eq1}). The magnon-photon coupling strength and the magnon Kerr coefficient are respectively denoted by $g_1=g$ and $K_1=K$. Our analysis focuses on the equal-time second-order correlation function of the photons in the cavity. The
    Fock-state basis of the system is denote by $\left| {nm} \right\rangle  = \left| n \right\rangle  \otimes \left| m \right\rangle $, with $n$ being the number of photons in the microwave cavity and $m$ the number of magnon. 
    In the weak pumping regime, $\Omega/\kappa_{c(m)}\ll1$, the photon number is small, so we can work within the few-photon subspace spanned by the basis states $|0\rangle_c$, $|1\rangle_c$, and $|2\rangle_c$. Therefore, the state of the system at arbitrary time can be expressed as
    \begin{align}
    	| \psi_t \rangle  =& C_{00}|0\rangle_c|0\rangle_m  + C_{10}|1\rangle_c|0\rangle_m  + C_{01}|0\rangle_c|1\rangle_m    \notag\\ 
    	&+ C_{20}|2\rangle_c|0\rangle_m+ C_{11}|1\rangle_c|1\rangle_m  + C_{02}|0\rangle_c|2\rangle_m,\label{eq4} 
    \end{align}
   where $C_{ij}$ with $i,j=0,1,2$ are the probability amplitudes. By substituting the state $|\psi_t\rangle$ into the Schr\"{o}dinger equation, the following equations of motion for the probability amplitudes can be obtained,
    \begin{align}\label{q7}
    			i{{\dot C}_{00}} =& \Omega{C_{10}}, \notag\\ 
    			i{{\dot C}_{10}} =& \Delta_c^\prime C_{10} + g{C_{01}} + \sqrt {2} \Omega {C_{20}} + \Omega{C_{00}}, \notag\\ 
    			i{{\dot C}_{01}} =& g{C_{10}} + (\Delta_m^\prime+K){C_{01}} + \Omega{C_{11}}, \notag\\ 
    			i{{\dot C}_{20}} =& 2\Delta_c^\prime{C_{20}} + \sqrt {2} \Omega{C_{10}}  +\sqrt {2} g {C_{11}},\\
    			i{{\dot C}_{11}} =& \Omega{C_{01}}  +\sqrt {2} g ({C_{20}} + {C_{02}}) +(\Delta_c^\prime+\Delta_m^\prime +K) {C_{11}},\notag\\ 
    			i{{\dot C}_{02}} =& \sqrt {2} g{C_{11}} +2(\Delta_m^\prime +2K) {C_{02}},\notag
    \end{align}
    where $\Delta_{c(m)}^\prime=\Delta_{c(m)}-i\kappa_{c(m)}/2$. In the long-time limit, the probalitity amplitudes can be attained by directly solving $\dot{C}_{ij}=0$.
    
    \begin{figure}
    	\includegraphics[scale=0.19]{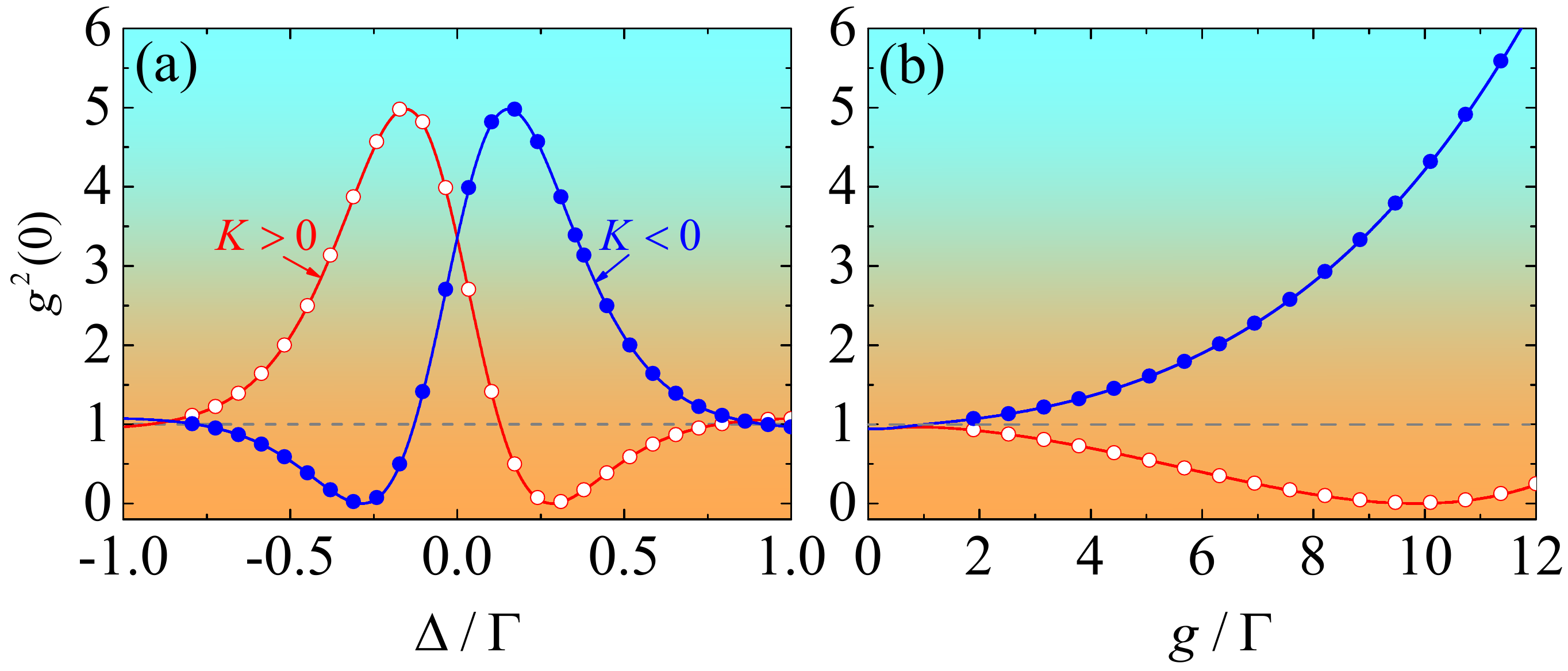}
    	\caption{${g^2}(0)$ versus the normalized (a) detuning $\Delta$ and (b) magnon-photon coupling strength $g$. The red (blue) curve corresponds to the case of $K>0~(K<0)$. In (a), $g=g_{\rm opt}=9.88\Gamma$, and in (b), $\Delta=\Delta_{\rm opt}=0.287\Gamma$. Other parameters are $\Gamma/2\pi=1$ MHz, $|K|/\Gamma=4\times 10^{-3}$, and $\Omega/\Gamma=0.1$.}  \label{fig2}
    \end{figure}
    When the system is in the state (\ref{eq4}), the equal-time second-order correlation function of the photons can be calculated as
    \begin{align}
    	g^2(0)\equiv\frac{\langle c^\dag c^\dag c c\rangle}{\langle c^\dag c\rangle^2}=\frac{2|C_{20}|^2}{(|C_{10}|^2+|C_{11}|^2+2|C_{20}|^2)^2}.\label{eq7} 
    \end{align}
    In the weak pumping regime ($\Omega\ll\Gamma$), we have $|C_{10}|^2\gg|C_{11}|^2,|C_{20}|^2$. This means that the probability of finding one photon in the cavity is much larger than that of simultanesouly finding one photon and one magnon, which is also much larger than that of finding two photons in the cavity. As a result, $g^2(0)\approx2|C_{20}|^2/|C_{10}|^4<1$, i.e., the photon blockade is achieved. Since the probabilities in Eq.~(\ref{eq7}) are affected by the magnon Kerr effect ($K$) [see Eq.~(\ref{q7})], the so-called nonreciprocal photon blockade can be achieved via changing the direction of the magnetic field (i.e., $K>0$ or $K<0$). To show this, we analytically plot the equal-time second-order correlation $g^2(0)$ versus the normalized detuning $\Delta/\Gamma$ and coupling strength $g/\Gamma$ in Fig.~\ref{fig2}, where $\omega_c=\omega_m=\omega$ (equivalently, $\Delta_c=\Delta_m=\Delta$) and $\kappa_c=\kappa_m=\Gamma$ are assumed for simplicity hereafter. The red (blue) curve denotes $K>0~(K<0)$, corresponding to the case that the magnetic field is aligned along the crystal axis $[100]~([110])$. From Fig.~\ref{fig2}(a), we show that the perfect photon blockade can be realized by tuning $\Delta$ when the magnon-photon coupling strength $g$ is fixed at its optimal value. For $K>0$ and $K<0$, the nonreciprocal photon blockade is predicted. When the positive optimal value of the detuning $\Delta_{\rm opt}/\Gamma=0.287$ is chosen [see Fig.~\ref{fig2}(b)], $g^2(0)$ decreases first from $g^2(0)=1$ to $g^2(0)=0$ and then increases with increasing $g$ when $K>0$. But when $K<0$, $g^2(0)$ monotonically increases, resulting in photon bunching [$g^2(0)>1$].  To demonstrate the validity of our approximate analysis, we also perform the numerical simulation by using the Lindblad master equation
    \begin{align}
    	\dot{\rho}=i[\rho,H_{\rm rf}]+\frac{\kappa_c}{2}\mathcal{L}[c]\rho+\frac{\kappa_m}{2}\mathcal{L}[m]\rho,
    \end{align}
    where $\rho$ is the density matrix of the considered system, $\mathcal{L}[o]\rho=2o\rho o^\dag-o^\dag o\rho-\rho o^\dag o$ is the Lindblad operator. Obviously, the analytical result well matches the simulation (see the circles and squares in Fig.~\ref{fig2}). The mechanism of the photon blockade can be explained by the destructive interference between two transition paths [see Fig.~\ref{fig1}(b)]. One path is formed by directly pumping the vaccum cavity to the cavity having two photons, i.e., $|0\rangle_c|0\rangle_m\rightarrow |1\rangle_c|0\rangle_m\rightarrow|2\rangle_c|0\rangle_m$. The other path is formed by the strong coupling between the magnons and photons. Specifically, when one photon is excited in the cavity, the magnon-photon coupling leads to the transition between the states $|1\rangle_c|0\rangle_m$ and $|0\rangle_c|1\rangle_m$. Then the pumpling field excites the state $|0\rangle_c|1\rangle_m$ to the state $|1\rangle_c|1\rangle_m$. Due to the photon-magnon coupling, the state $|1\rangle_c|1\rangle_m$ further transits to the states $|2\rangle_c|0\rangle_m$ and $|0\rangle_c|2\rangle_m$. During these transitions, the frequency shift induced by the magnon Kerr effect is positive (negative) for $K>0~(K<0)$. This indicates when the photon blockade is achieved at $K>0$ ($K<0$) for fixed parameters, the reversed photon bunching, i.e., $g^2(0)>1$, is predicted at $K<0$ ($K>0$), as demonstrated in Fig.~\ref{fig2}(a).    

\begin{figure}
    	\includegraphics[scale=0.285]{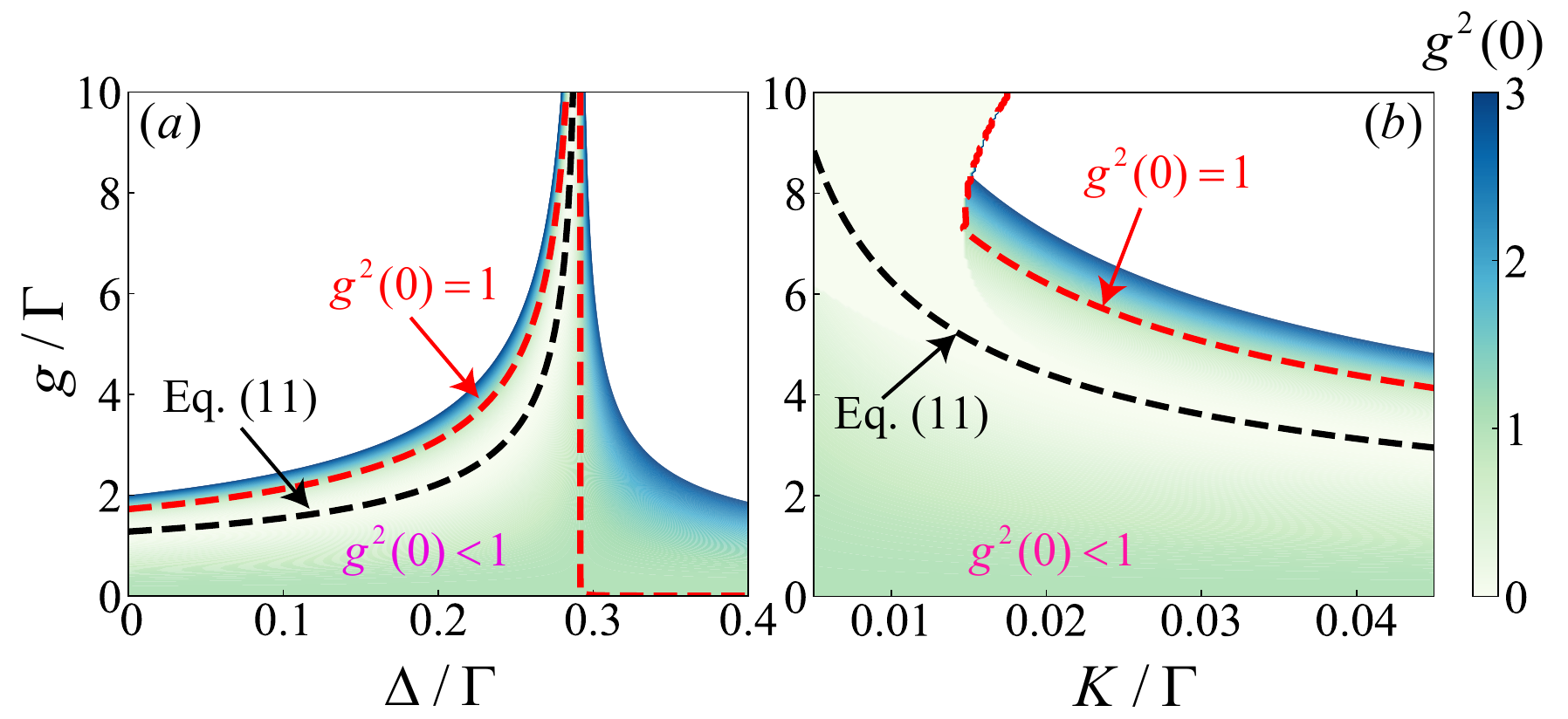}
    	\caption{(a) Density plot of $g^2(0)$ vs the normalized coupling strength $g/\Gamma$ and the normalized detuning $\Delta/\Gamma$. (b) Density plot of $g^2(0)$ vs the normalized coupling strength $g/\Gamma$ and the normalized Kerr coefficient $K/\Gamma$. In panels (a) and (b), the red dashed curve denotes $g^2(0)=1$, the black dashed curve satisfies the optimal conditon in Eq.~(\ref{eq11}), and the light green zone means photon blockade, i.e., $g^2(0)<1$. Other parameters are the same as those in  Fig.~\ref{fig2}.}\label{fig3}
    \end{figure}

    From Fig.~\ref{fig2}, one can find that the optimal coupling strength $g_{\rm opt}$ and frequency detuning $\Delta_{\rm opt}$ for a given $K$ must exist for prediction of the perfect photon blockade [${g^2}(0) =0$]. This indicates that the probability of simultaneously finding two photons in the cavity is nearly zero [see Eq.~(\ref{eq7})], i.e., $|C_{20}|^2\approx0$, which can be directly convinced by the simulation results in Fig.~\ref{fig3}.  To analytically obtain the optimal parameters, the perfect photon blockade condition can be specifically rewritten as
        \begin{align}
    	\frac{g^2K}{\Delta_m^\prime +2 K}+(\Delta_c^\prime+\Delta_m^\prime + K)(\Delta_m^\prime + K)= \Omega ^2,\label{eq8} 
    \end{align}   
    or equivalently,
    \begin{align}\label{eq9}
    	2\Omega^2+\Gamma^2=&12\Delta^2+28\Delta K+14 K^2,\notag\\
    	\frac{g^2K}{4\Delta+3K}=& (\Delta +2 K)^2+\Gamma^2/4.
    \end{align}
    From the second equality in Eq.~(\ref{eq9}), the inequality
    \begin{align}\label{eq10}
    	(4\Delta+3K)K>0
    \end{align}
    can be directly obtained for a given $g$. This means that the perfect photon blockade can only be predicted in the region of $\Delta>-3K/4~(<-3K/4)$ for $K>0~(<0)$. In addition, the optimal coupling strength 
    \begin{align}\label{eq11}
    	g_{\rm opt}=\sqrt{\frac{4\Delta_{\rm opt}+3K}{K}[(\Delta_{\rm opt} +2 K)^2+\Gamma^2/4]}
    \end{align}
is required to realize perfect photon blockade for a given $K$, where the optimal paramter $\Delta_{\rm opt}$ is given by the first equality in Eq.~(\ref{eq9}), i.e.,
    \begin{align}\label{eq12}
    \Delta_{\rm opt}=\frac{-7K\pm\sqrt{7K^2+6\Omega^2+3\Gamma^2}}{6}\approx\pm\frac{\sqrt{3}}{6}\Gamma.
    \end{align}
    The second approximate equality is established because $K,\Omega\ll\Gamma$ is taken. The sign '$+~(-)$' corresponds to $K>0~(<0)$.

    \begin{figure}
    	\includegraphics[scale=0.18]{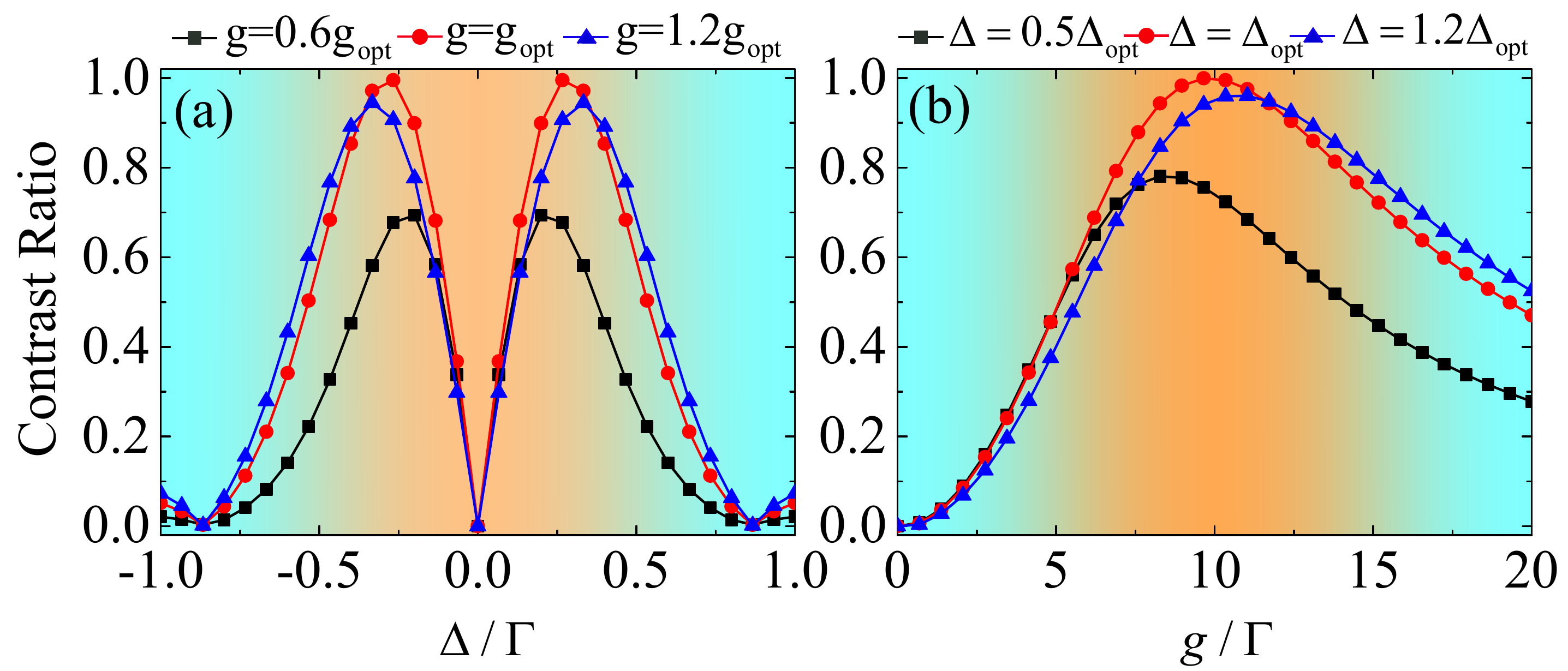}
    	\caption{(a) The contrast ratio $C$ versus the normalized detuning $\Delta$ with different magnon-photon coupling strengths $g=g_{\rm opt}=9.88\Gamma$ (red), $0.6g_{\rm opt}$ (black), and $1.2g_{\rm opt}$ (blue). (b) The contrast ratio $C$ versus the normalized coupling strength $g$ with different detunings $\Delta=\Delta_{\rm opt}=0.287\Gamma$ (red), $0.5\Delta_{\rm opt}$ (black), and $1.2\Delta_{\rm opt}$ (blue). Other parameters are the same as those in  Fig.~\ref{fig2}.}\label{fig4}
    \end{figure}
    
    To quantitatively charaterize the nonreciprocal photon blockade, a bidirectional contrast ratio is introduced, i.e.,
    \begin{align}
    	C= \left| {\frac{{g_{K > 0}^2(0) - g_{K < 0}^2(0)}}{{g_{K > 0}^2(0) + g_{K < 0}^2(0)}}} \right|\in[0,1],
    \end{align}
    where $C=1~(0)$ denotes the ideal nonreciprocal (reciprocal) photon blockade. The larger the contrast ratio $C$, the stronger the nonreciprocity of the photon blockade. In Fig.~\ref{fig4}(a), we show the behavior of the contrast ratio with the normalized detuning $\Delta/\Gamma$ with different magnon-photon couplings. Clearly, the nonreciprocity and reciprocity for the photon blockade can be switched by tuning the detuning $\Delta$. When the coupling strenth is optimal (i.e., $g_{\rm opt}=9.88\Gamma$), the ideal nonreciprocal photon blockade can be attained. But when the coupling strength deviates from the optimal value such as $g=0.6g_{\rm opt}$ and $g=1.2g_{\rm opt}$, the maximum nonreciprocity of the photon blockade has a different degree of reduction~(see curves marked by squares and triangles). In Fig.~\ref{fig4}(b), we also investigate the contrast ratio with the normalized coupling strength $g/\Gamma$ with different detunings. By increasing $g$, one can see that the contrast ratio increases first to its maximum and then decreases. At the optimal detuning $\Delta_{\rm opt}=0.287\Gamma$, the ideal nonreciprocal photon blockade ($C=1$) is predicted. Deviating from this optimal value such as $\Delta=0.5\Delta_{\rm opt}$ (the black curve with squares) and $\Delta=1.2\Delta_{\rm opt}$ (the blue curve with triangles), the maximum nonreciprocity of the photon blockade reduces.

    \section{Nonreciprocal photon blockade with Two spheres}\label{s4} 
     
     We next study the photon blockade in the considered system consisting of two YIG spheres simultanesouly coupled to a cavity [see Eq.~(\ref{eq3})  with $j=2$]. For conveniance, we define two parameters $\zeta_g=g_1/g_2$ and $\zeta_K=K_1/K_2$ as the relative coupling strength and Kerr coefficient, respectively. In the weak pumping regime, the state of the considered system governed by Eq.~(\ref{eq3}) with $j=2$ becomes	
	\begin{align}\label{eq14}
		|\psi_t^\prime\rangle =&C_{000}|000\rangle+C_{001}|001\rangle+C_{100}|100\rangle +C_{010}|010\rangle\notag\\
		+&C_{200}|200\rangle+C_{110}|110\rangle+C_{011}|011\rangle+C_{101}|101\rangle\notag\\
		+&C_{020}|020\rangle+C_{002}|002\rangle
	\end{align}
	when the system is initially prepared in the state $|0\rangle_c|0\rangle_1|0\rangle_2$, where the subscripts $c$, $1$ and $2$ denote the cavity mode, the spheres $1$ and $2$. Following the procedure of calculating the equal-time second-order correlation fuction in the case of a single sphere, we have  
    \begin{align}
		G^2(0) &=\frac{2|C_{200}|^2}{(|C_{100}|^2 + |C_{110}|^2 + |C_{101}|^2 + 2|C_{200}|^2)^2},\label{eq15}
	\end{align}
    which charaterizes the equal-time second-order correlation fuction in the presence of two YIG spheres. Here we have used the approximation $|C_{200}|^2\ll|C_{110}|^2,|C_{101}|^2\ll |C_{100}|^2$. This directly leads to photon blockade, i.e., $G^2(0)<1$. To address this, we further consider the following two scenarios: (i) The directions of two magnetic fields are identical ($\zeta_K>0$); (ii) the directions of two magnetic fields are opposite ($\zeta_K<0$). When $\zeta_K>0$,  the predicted nonreciprocal photon blockade is similar to that of a single sphere (see Fig.~\ref{fig2}), which has been numerically checked. Therefore, we do not provide discussions here anymore. 
       \begin{figure}
    	\includegraphics[scale=0.185]{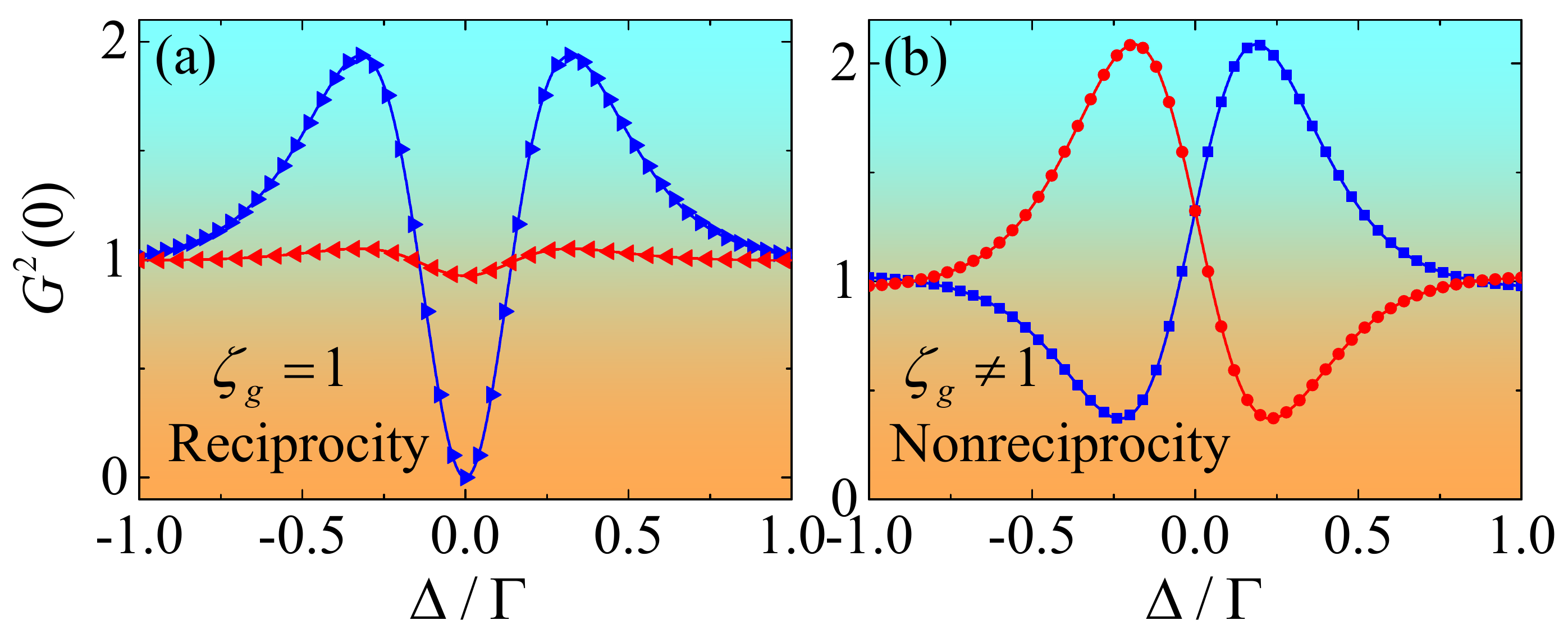}
    	\caption{$G^2(0)$ versus the normalized detuning $\Delta$ with (a) $g_1=g_2$ and (b) $g_1\neq g_2$ in the case of $\zeta_K<0$. In (a), the red (blue) curve corresponds to $g_1=g_2=15\Gamma$ ($=g_{\rm opt}=63\Gamma$). Other parameters are the same as those in Fig.~\ref{fig2}.}\label{fig5}
    \end{figure}

    Interestingly, the situation of $\zeta_K<0$ is completely different from that of $\zeta_K>0$. For simplicity, we assume that the magnons in two spheres have the same absolute values, i.e., $|\zeta_K|=1$, equivalently $|K_1|=|K_2|$. In the following discussion, we label the scenario of $K_1>0$ and $K_2<0$  ($K_1<0$ and $K_2>0$) as $K_{+-}$ ($K_{-+}$). When the magnons in two YIG spheres are identically coupled to the cavity ($\zeta_g=1$), only the reciprocal photon blockade is predicted for $K_{+-}$ and $K_{-+}$ [see red or blue curve in Fig.~\ref{fig5}(a)]. This is due to the fact that the transitions $|000\rangle\rightarrow|100\rangle\rightarrow|001\rangle\rightarrow|101\rangle\rightarrow|200\rangle$ and $|000\rangle\rightarrow|100\rangle\rightarrow|010\rangle\rightarrow|110\rangle\rightarrow|200\rangle$ play the same role in destructively interfering with the transition $|000\rangle\rightarrow|100\rangle\rightarrow|200\rangle$ when $\zeta_g=1$ and $|\zeta_K|=1$ [see Fig.~\ref{fig1}(c)]. To obtain a {\it visible} photon blockade [$G^2(0)\ll1$], the large magnon-coupling strengths are needed. At the optimal coupling strength $g_{\rm opt}=63\Gamma$, we find that the perfect photon blockade is achieved at $\Delta_{\rm opt}=0$, as shown by the blue curve in Fig.~\ref{fig5}(a). This optimal coupling strength can be experimentally realized owing to the achieved strong and ultra-strong photon-magnon interactions~\cite{Zhang-2014,Justin-2019,Kostylev-2016}. However, when $\zeta_g\neq1$ (i.e., $g_1\neq g_2$), the nonreciprocal photon blockade is clearly observed [see Fig.~\ref{fig5}(b)], where the red (blue) curve corresponds to $K_{+-}$ ($K_{-+}$). To realize this nonreciprocal photon blockade, the required magnon coupling strength is relatively smaller than that of $\zeta_g=1$. This means that the nonreciprocal photom blockade can be engineered by using the asymmetric and relatively small magnon-photon coupling strength, making the proposal more feasible in experiments. At $\Delta/\Gamma=+2.87$ ($-2.87$), the optimal photon blockade occurs for $K_{+-}$ ($K_{-+}$). The mechanism of the nonreciprocal photon blockade at $\zeta_g\neq1$ can be interpreted as follows: For $K_{+-}$ [see the red levels in Fig.~\ref{fig1}(c)], the transition $|000\rangle\rightarrow|100\rangle\rightarrow|010\rangle\rightarrow|110\rangle\rightarrow|200\rangle$ is allowed at $\Delta>0$, while the transtion  $|000\rangle\rightarrow|100\rangle\rightarrow|001\rangle\rightarrow|101\rangle\rightarrow|200\rangle$ is forbidden due to the Kerr effect induced large detuning. As a result, the photon blockade is caused by the destructive interference between the former and the direct pumping path $|000\rangle\rightarrow|100\rangle\rightarrow|200\rangle$. On the contrary, the transition $|000\rangle\rightarrow|100\rangle\rightarrow|010\rangle\rightarrow|110\rangle\rightarrow|200\rangle$ is forbidden at $\Delta<0$, while the transtion $|000\rangle\rightarrow|100\rangle\rightarrow|001\rangle\rightarrow|101\rangle\rightarrow|200\rangle$ is allowed, giving rise to photon blockade for $K_{-+}$ [see the blue levels in Fig.~\ref{fig1}(c)].
     
     \begin{figure}
    	\includegraphics[scale=0.18]{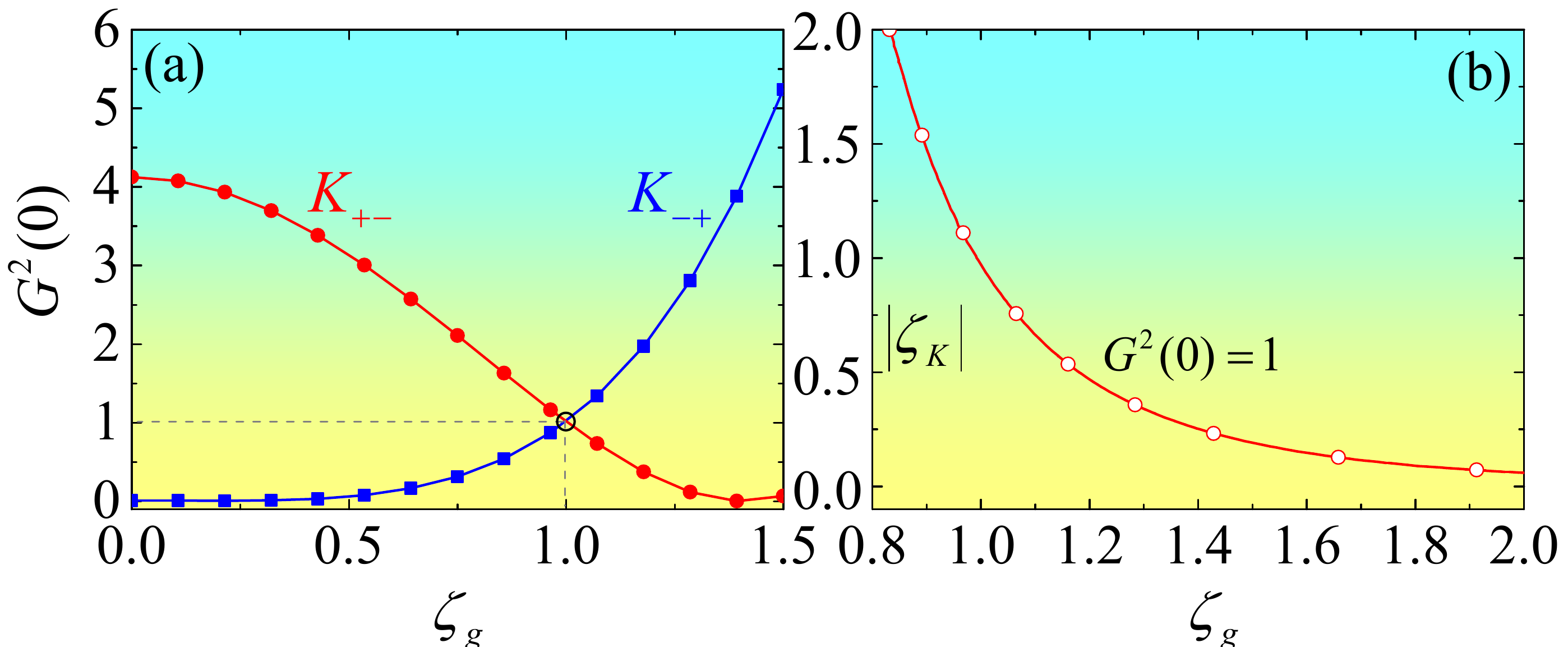}
    	\caption{(a) $G^2(0)$ versus the relative coupling strength $\zeta_g$ with $\Delta=-0.287\Gamma$ in the case of $\zeta_K<0$, where the red (blue) curve corresponds to $K_{+-}~(K_{-+})$. (b) The contourplot of $G^2(0)=1$ versus $\zeta_g$ and $|\zeta_K|$. Other parameters are the same as those in Fig.~\ref{fig2}.}\label{fig6}
    \end{figure}
    Figure~\ref{fig6}(a) further examines the behavior of the photon blockade with the relative coupling strength $\zeta_g$, where $g_2/\Gamma=9.88$ is fixed. In the absence of one sphere such as the sphere 1 ($\zeta_g=0$), the photons in the cavity is bunching (antibunching) for the case of $K_{+-}$ ($K_-+$). By coupling the sphere 1 to the cavity and continuously increasing $g_1$, we find that the property of the statistic photons are changed from bunching to antibunching (blockade) for $K_{+-}$, and conversely, from antibunching to bunching for $K_{-+}$. This indicates that the nonreciprocal photon blockade can be achieved in a broad range of the parameter $\zeta_g$. Note that at $\zeta_g=1$ ($g_1=g_2=9.88\Gamma$), $G^2(0)=1$ for both $K_{+-}$ and $K_{-+}$ (see the crosspoint), meaning that the nonreciprocity disappears and photons satisfies Poissionian distribution. Figure~\ref{fig6}(b) reveals the relationship between $\zeta_g$ and $|\zeta_K|$ when $G^2(0)=1$. With increasing $\zeta_g$, the relative Kerr coefficient decreases. This suggests that the crosspoint in Fig.~\ref{fig6}(a) has a right (left) shift with increasing (decreasing) $\zeta_g$ when $|\zeta_K|<1~(>1)$.
    
       \begin{figure}
    	\includegraphics[scale=0.18]{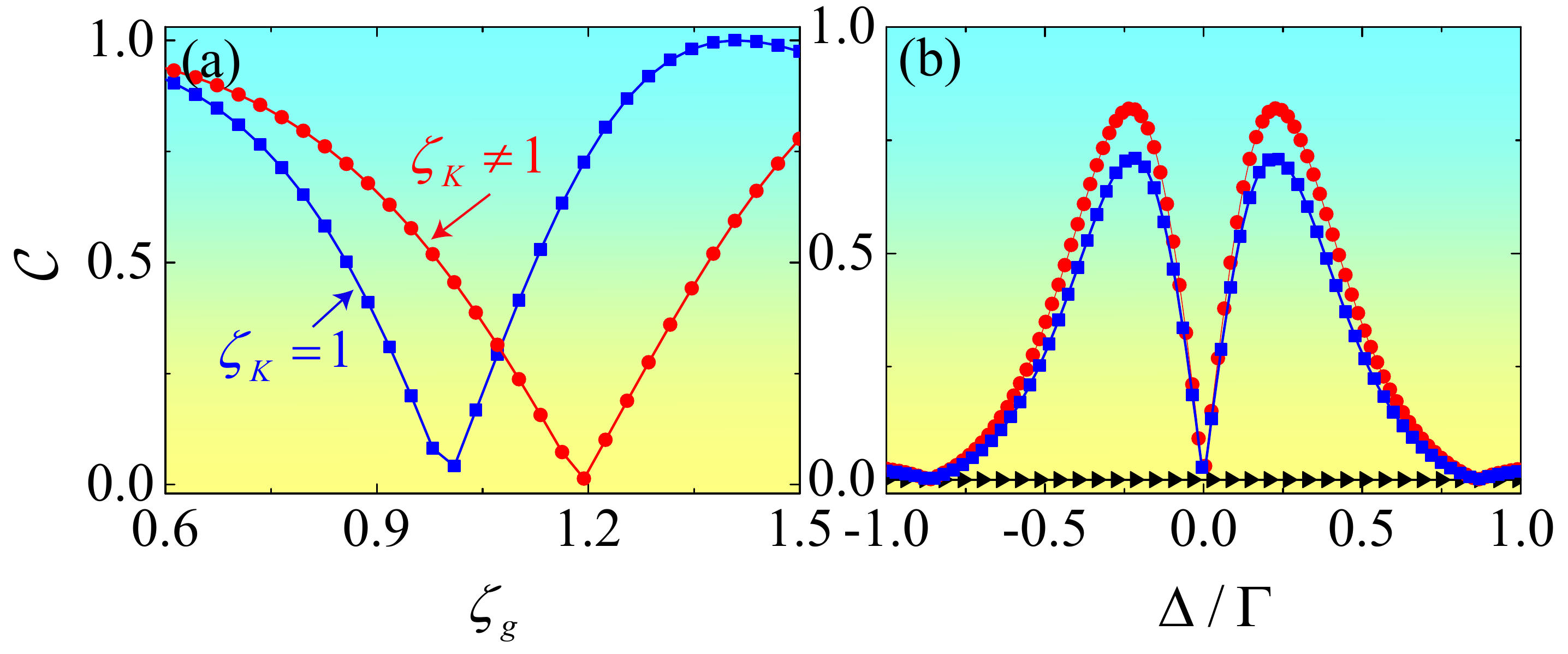}
    	\caption{The contrast ratio $\mathcal{C}$ versus (a) the relative coupling strength $\zeta_g$ and (b) the normalized detuning $\Delta$. In (a), $\zeta_K=1$ with $|K_1|=|K_2|=4\times10^{-3}\Gamma$ (blue) and $\zeta_g\neq1$ with $|K_2|=2|K_1|=4\times10^{-3}\Gamma$ (red).  In (b), the black curve denotes $g_1=g_2=9.88\Gamma$ and $|K_1|=|K_2|=4\times10^{-3}$, the red curve denotes $g_1/\Gamma=12,~g_2/\Gamma=9.88$, and $|K_1|=|K_2|=4\times10^{-3}$, the blue curve denotes $g_1=g_2=9.88\Gamma$ and $|K_2|=4|K_1|=4\times10^{-3}$. Other parameters are the same as those in Fig.~\ref{fig2}.}\label{fig7}
    \end{figure}
    \begin{figure}
    	\includegraphics[scale=0.18]{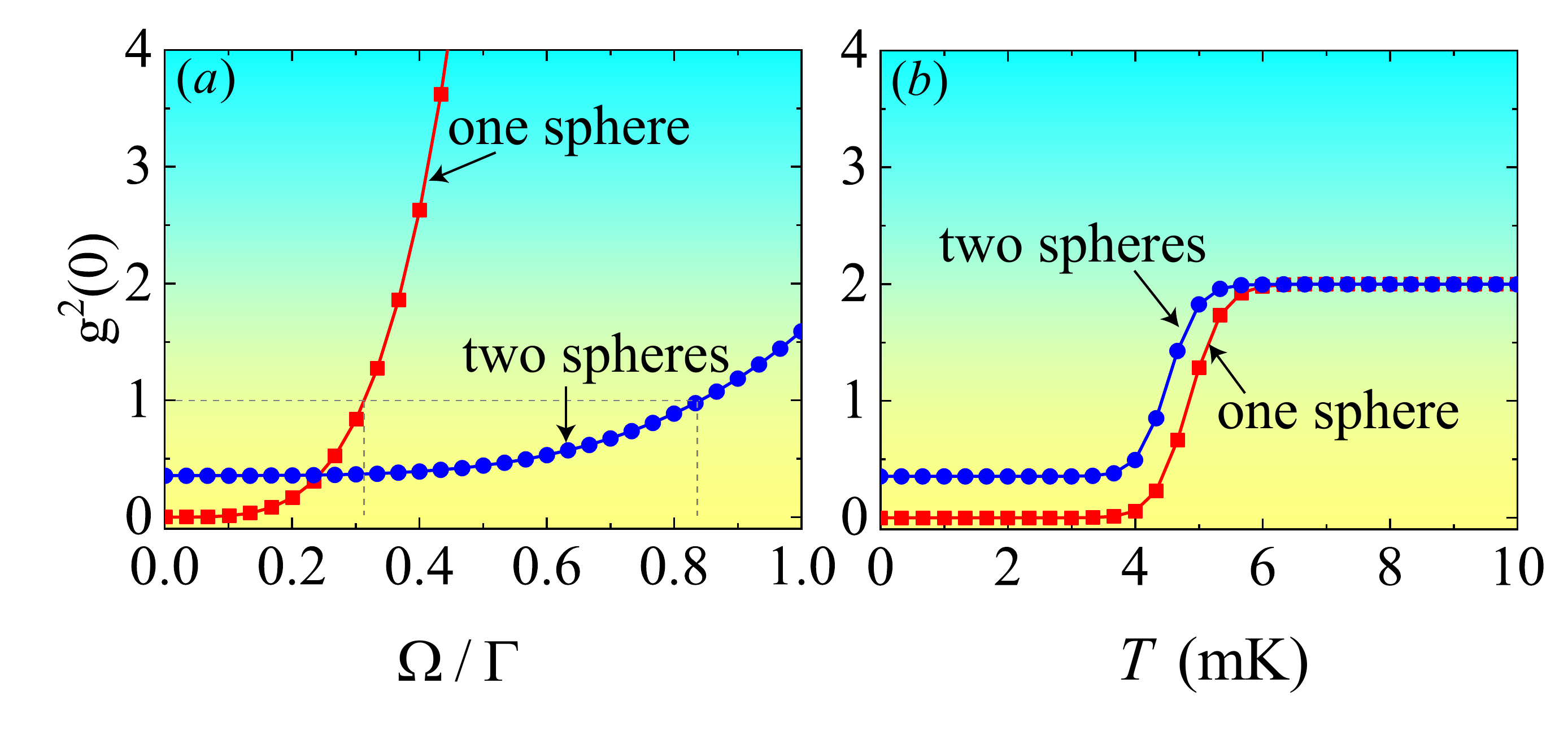}
    	\caption{The second-order correlation function $g^2(0)$ vs (a) the normalized Rabi frequency of the pumping field and (b) the bath temperature, where the red (blue) curve denotes the case of the single sphere (two spheres). Other parameters are the same as those in Fig.~\ref{fig2}.}\label{fig8}
    \end{figure}

    To describe the nonreciprocity of the photon blockade induced by the opposite Kerr effects of the magnons in two spheres, a contrast ratio $\mathcal{C}$ is defined as
    \begin{align}
    	\mathcal{C}=\left| {\frac{{G_{K_{+-}}^2(0) - G_{K_{-+}}^2(0)}}{{G_{K_{+-}}^2(0)+ G_{K_{-+}}^2(0)}}} \right|.
    \end{align}
    In Fig.~\ref{fig7}, we respectively plot it versus the relative coupling strength $\zeta_g$ and the normalized detuning $\Delta/\Gamma$. One can see that the nonreciprocity can be well tuned between $0$ (reciprocity) and $1$ (nonreciprocity) by the relative coupling strength $\zeta_g$ in Fig.~\ref{fig7}(a) when $|\zeta_K|=1$. In particular, the nonreciprocity disappears at $\zeta_g=1$, consistent with above discussions. To recover the nonreciprocity, aysmmetric coupling strengths ($\zeta_g\neq1$) or Kerr coefficents ($|\zeta_K|\neq 1$) can be employed, as demonstrated by the the blue and red curves, respectively. Obviously, the nonreciprocity of the photon blockade can also be controlled by $\zeta_g$ for the asymmetric Kerr coefficents ($|\zeta_K|\neq 1$). When the magnon-photon coupling strenghts are fixed, the contrast ratio can be tuned by the normalized detuning $\Delta/\Gamma$ in Fig.~\ref{fig7}(b). Specifically, only reciprocal photon blockade is predicted ($\mathcal{C}=0$)  at $\zeta_g=1$,  $|\zeta_K|=1$ (see the black curve). However, one of the conditions is broken, i.e., $\zeta_g\neq 1$ and  $|\zeta_K|=1$, or $\zeta_g=1$ and $|\zeta_K|\neq1$, the nonreciprocity of the photon blockade can be observed.

	\section{Discussion and conclusion}\label{s6}
	
	 Before concluding, we give a brief study of the Rabi frequency of the weak pumping field and the effect of the bath temperature on the photon blockade. From Fig.~\ref{fig8}(a), one can see that the photon blockade can be realized at $\Omega<0.31\Gamma$ ($\Omega<0.84\Gamma$) in the presence of single YIG sphere (two YIG spheres). This indicates that the range of $\Omega$ for achieving the photon blockade can be widened via increasing the number of YIG spheres. Figure~\ref{fig8}(b) shows the impact of the bath temperature on the photon blockade. Obviously, $g^2(0)$ is nearly unchanged when $T<4$ mK for both the cases of single sphere and two spheres. But when the temperature crosses the point $T=4$ mK, $g^2(0)$ has a sudden increase. For the case of single sphere (two spheres), photon blockade disappears when $T>4.45$ mK ($T>4.5$ mK). It is also evident that the proposed system including one sphere can have better photon blockade effect than the case of two spheres at a certain temperature. 
     
     In summary, we have proposed a nonlinear cavity-magnon system to study the nonreciprocal photon blockade. The nonreciprocity stems from the direction-dependent Kerr effect of magnons in the YIG sphere. For a single sphere case, the nonreciprocal destructive interference between two paths leads to nonreciprocal photon blockade by varying the Kerr coefficient from positive to negative (or vice versa). By optimizing the system parameters, perfect nonreciprocal photon blockade can be predicted and finely tuned. For the case of two spheres with opposite Kerr coefficients, only reciprocal photon blockade can be predicted when two cavity-magnon coupling strengths and Kerr coefficients are symmetric. However, when two coupling strengths or Kerr coefficients becomes asymmetric, nonreciprocal photon blockade appears. This indicates that  the transition between reciprocity and nonreciprocity of photon blockade can be arbitrarily switched in a two-sphere cavity-magnon system. Our study paves a potential way to engineer nonreciprocal devices in nonlinear cavity magnonics.

	
	This work was supported by Zhejiang Provincial Natural Science Foundation of China under Grant No.~LY24A040004, the National Natural Science Foundation of China under Grant No. 11804074, and the Natural Science Foundation of Hubei Province of China under Grant No.~2022CFB509.


\end{document}